\documentclass[9pt]{article}
\usepackage{graphicx}
\usepackage{bookman}
\usepackage{array}
\usepackage{latexsym}
\usepackage{amssymb}
\usepackage{amsbsy}
\usepackage{cite}
\usepackage[T1]{fontenc}
\begin{document}
\title{Observational consequences of a dark interaction model}
\author{M.de Campos$^{(1)}$\\
}\maketitle
\footnote[1]{Physics Department - Roraima Federal University.  Av. Ene Garcez 2413, Campus do Paricar\~ana, Bloco V,
Bairro Aeroporto, Boa Vista, Roraima, Brasil.69304-000,email:campos@if.uff.br.}
\begin{abstract}
We study a model with decay of dark energy and creation of the dark matter particles. We integrate
 the field equations and find the transition redshift where the evolution process of the universe change 
the accelerated expansion, and discuss the luminosity distance, acoustic oscillations and the statefinder parameters.
\end{abstract}
\section{Introduction}             
Before  the   results  from  supernova of the type IA observations   appear  in  the
literature, that  indicates an accelerated expansion  of the universe,
L. Krauss  and M.  Turner  have called our attention that  ``The Cosmological
Constant  is  Back  ''.  They  cited  the  age  of the  universe,  the
formation of  large scale structure and the  matter content of the
universe as the data that  indicates the insertion of  cosmological
constant \cite{Krauss}.
The  cosmic   microwave  background  radiation
anisotropy and large scale structure, also indicates this acceleration
expansion of the universe    \cite{Riess}- \cite{Spergel}.  Besides, the analysis of 158 SNe realized by 
Riess et al. \cite{Riess1} point out the present acceleration ($q<0$) at 99.2$\%$ at confidence level. 

The mechanism that triggered the  acceleration of the universe has not
been identified, and the simplest  explanation for  this process is
the inclusion  of a non null cosmological  constant.
However,  the  inclusion  of   cosmological  constant  creates  new
problems. Some of them are  old, as the discrepancy among the observed
value  for the  energy  density of  the  vacuum and  the large  value
suggested   by   the    particle   physics   models   \cite{Weinberg},
\cite{Garrig}.  In spite of  the problems caused by the inclusion of
$\Lambda $, the cosmological scenario with $\Lambda$  has a good agreement with respect
to  the estimate age  of the  universe, the  anisotropy of  the microwave
background radiation and the supernova experiments. Besides, making
several assumptions concerning with the spectrum of fluctuations in the 
early  universe and  the formation of the galaxies,  G. Efstathiou suggests 
that the small value  of  cosmological constant  can be explained by 
the anthropic principle \cite{Efstathiou}.  Although the inclusion of $\Lambda$ is the simplest explanation for 
the cosmic acceleration, there are a lot of alternatives to explain the accelerated expansion.  See the reviews 
\cite{Carrol} and \cite{Peebles}, and the references therein.  So, we have experimental evidence for two extra 
components for the universe.  One is responsible for the cosmic acceleration and represents about 70$\%$ of material
 content, and the other acts gravitationally as ordinary matter, but is not baryonic.

The evidence that these new components of the universe, dark matter and dark energy are different substances
 has been considered in the literature \cite{Sandvik}.  Generally, the dark  matter component is considered as
 weakly interacting massive particles 
and the dark energy component is associated to some form of a scalar field.
A link between both components to a scalar field is studied by Padmanabhan and Choudhury 
\cite{Padmanabhan}.  Although, today, both  components are  unknown  is respect  to the  your
nature.

An alternative  model that  furnish a negative pressure in  the cosmic fluid and
results in an  accelerated expansion of the universe  is known as open
system cosmology (OSC) \cite{Prigogine}.  In OSC model the particle number
in  the universe  is not conserved and  the energy-momentum  tensor is
reinterpreted  in the  Einstein's field equations, where appear an extra negative pressure known as creation 
pressure \cite{Lima1},\cite{Lima2}.
The creation process is due to the expenses of the gravitational field
and is an irreversible process.  One of the attractive features of the
hypothesis of  particle production in OSC model is  the relation among  the large
scale   properties   of  the   universe   and   the  atomic   phenomena
\cite{McCrea}.

The coupling into dark matter and dark energy has been considered within three possibilities in literature.
First, considering the dark  matter decaying in dark energy; second, the dark energy decaying to dark matter;
 or interaction in both directions. See  \cite{Pereira} and references therein for examples for every one of the alternatives. 

On the other hand, the second law of thermodynamic favored the second possibility \cite{Wang}, \cite{Pavon}.
 Consequently, if each component is not conserved individually, the chemical potential  of at least one of 
the dark components is not null \cite{Pereira}, differently that appear in \cite{Wang}, \cite{Pavon} where
 both chemical potentials are zero. 

In this work we consider a different rate for diluting of the material components due to decaying of the 
dark energy into dark particles, as the same way that the authors considered in \cite{Meng}. Several aspects of 
this approach are investigate in \cite{Lima3}.  We study the transition of the accelerated expansion of the universe, 
the luminosity distance and the acoustic scale of the anisotropies of CMB, obtaining a validity interval for the parameter 
that governs the interaction between the dark components.  We finish this study with the statefinder pair $\{r,s\}$ that
 indicates the proximity of this model with LCDM model. We hope that in the future the statefinder parameter to be useful 
tools in testing interacting cosmologies.
\section{The cosmological model}
We consider the space-time as homogeneous and isotropic, characterized by the FRW metric
     \begin{equation}
       ds^2=dt ^2-R(t)^2[dr^2+r^2d\theta^2+r^2\sin{\theta}^2]d\phi ^2\, ,
        \end{equation}
and the energy momentum tensor as the usual perfect fluid, given by
   \begin{equation}
         T_{\mu \nu}= (\rho_{dm} + \rho_{de}+P)u_{\mu}u_{\nu}-P g_{\mu \nu}.
               \end{equation}
$P=P_{de}+P_{dm}$ is the total pressure, $\rho_{de}$ is the dark energy density, $\rho_{dm}$ is the dark matter density,
 while $u_\mu$ is the four velocity. Taking into account that the reference system is just the matter filling it, 
the field equations assumes the form
\begin{eqnarray}
\frac{\ddot R}{R}&=&-\frac{4}{3}\pi G (\rho_{dm}+\rho_{de}+3P)\, , \\
\frac{\dot R ^2}{R^2}&=& \frac{8\pi G}{3} (\rho_{dm}+\rho_{de})\, ,
\end{eqnarray}
where the spatial flatness is assumed, in accord with the data from WMAP \cite{Spergel1}.

Writing the conservation law as
\begin{equation}
u_{\mu}T^{\mu \nu} _{;\nu}=-u_{\mu}(\rho_{de} g^{\mu \nu})_{;\nu}\, ,
\end{equation}
its assumes the form
\begin{equation}
\dot \rho + 3\frac{\dot R}{R}\rho =-\dot \rho_{de}\, ,
\end{equation}
where $\rho = \rho_{dm}+ \rho_{de}$.

Although the vacuum component decays, we consider that the state equation remains with the usual state equation 
expression, $P_{de}=-\rho_{de}$. Some details about $\Lambda$ decaying model can be view in \cite{Carvalho}, and 
the thermodynamic behavior in \cite{Lima4}.

Considering the creation process, the dark matter density will dilute in a different rate, namely
\begin{equation}
\rho_{dm}=\rho_{dm0}R^{\epsilon -3}\, ,
\end{equation}
where the positive constant $\epsilon$ furnish the deviation from the process without decaying of the dark energy component,
and the subscript $0$ indicates the present time.

Rewritten Eq. (6) as 
\begin{equation}
\frac{d\rho_{dm}}{dR}+3\frac{\rho_{dm}}{R}=-\frac{d\rho_{de}}{dR}\, ,
\end{equation}
the integration results
\begin{equation}
\rho_{de}=\rho_{de0}-\frac{\epsilon}{3-\epsilon}\rho_{dm0}[1-R^{\epsilon -3}].
\end{equation}
With auxilious of Eq. (4), (7) and (9) we find the field equation
\begin{equation}
\dot R ^2-K_IR^2- K_{II}R^{\epsilon -1}=0\, ,
\end{equation}
where $K_I=\frac{8\pi G}{3}(\rho_{de0}-\frac{\epsilon \rho_{dm0}}{3-\epsilon})$ and 
$K_{II}=\frac{8\pi G}{3-\epsilon}\rho_{dm0}$, and the solution is given by
\begin{equation}
R(t)=(\frac{K_{II}}{K_I})^{\frac{1}{3-\epsilon}}\{\sinh{\sqrt{K_I}\frac{3-\epsilon}{2}t}\}^{\frac{2}{3-\epsilon}}\, .
\end{equation}
Consequently, the Hubble function and the deceleration parameter, as functions of the red-shift, are respectively:
\begin{eqnarray}
H(z)&=& H_0 \{\frac{3\Omega_{dm0}}{3-\epsilon}[(1+z)^{3-\epsilon}-1]+1\}^{\frac{1}{2}}\, ,\\
q(z)&=&-1+\frac{3-\epsilon}{2}\{(1+\frac{3\Omega_{de}-\epsilon}{3\Omega_{dm}})(1+z)^{\epsilon-3}\}^{-1} .
\end{eqnarray} 
The profiles for the Hubble function and deceleration parameter appear in the Fig.1 and Fig.2, respectively. 
\begin{figure}[!ht]
{\centerline{\includegraphics[width=6cm]{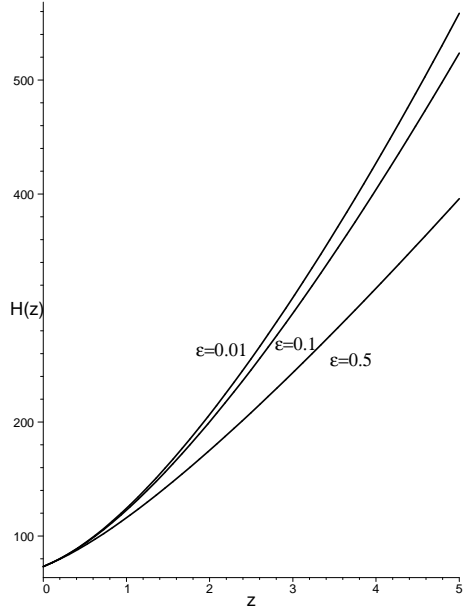}}}
\caption{ Profile for $H(z)\times z$.The values for $\epsilon$ considered are $\epsilon=0.01,\, 0.1,\, 0.5$. 
Note that the growing value for the $\epsilon$-parameter favored an older universe.}
\label{ Fig.1}
\end{figure}
\begin{figure}[!ht]
{\centerline{\includegraphics[width=6cm]{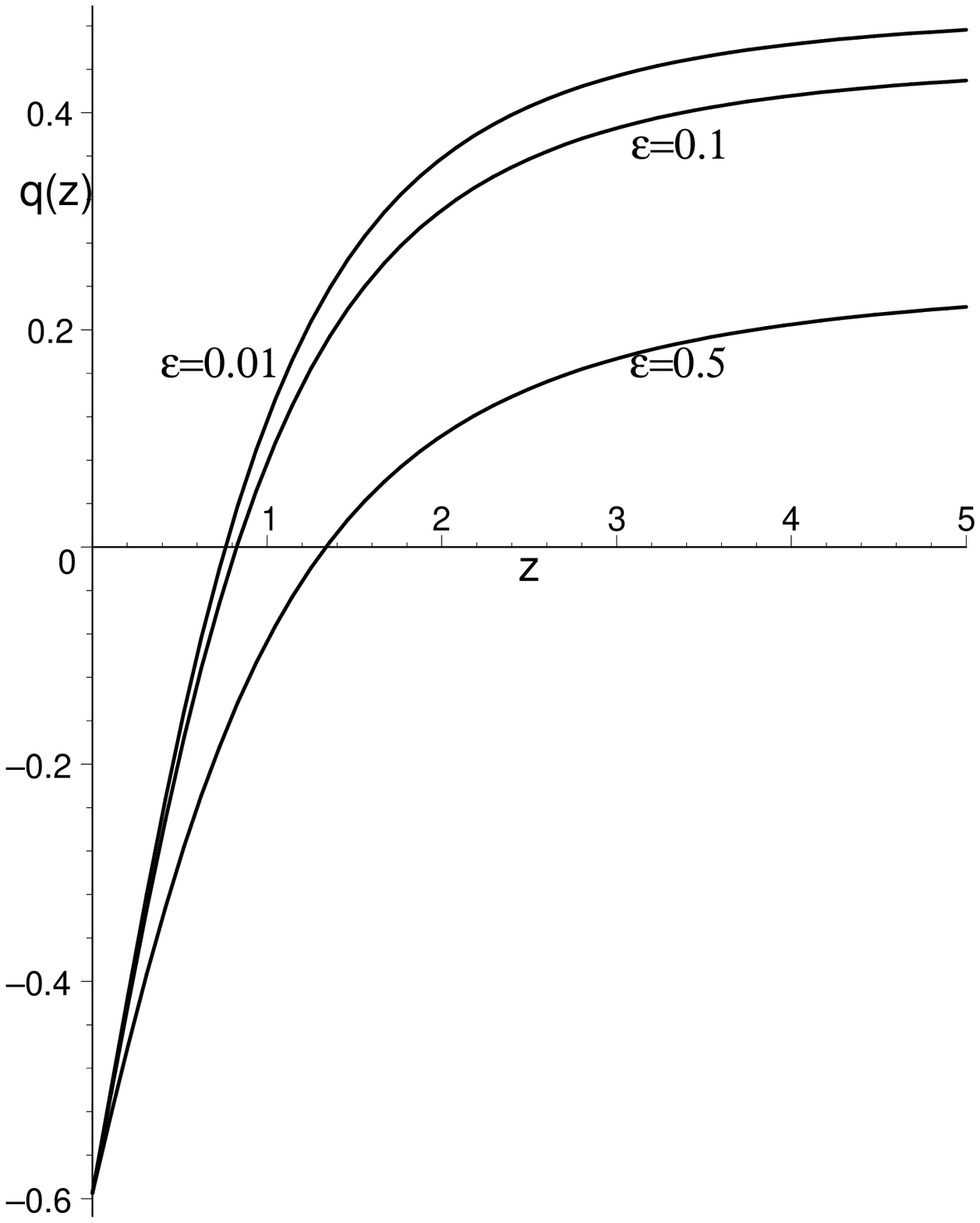}}}
\caption{ Profile for deceleration parameter as a function of the redshift. Note that the growing value for $\epsilon$ results
a highest redshift transition. Around the present time, we note by the
 graph, that the value for the deceleration parameter is $q_0=-0.59$, inside the interval established in 
\cite{Virey} for the present deceleration parameter, namely $q_0=-0.74 \pm 0.18$.}
\label{ Fig.2 }
\end{figure}

The age of the universe is one of the observational arguments for the existence of dark components \cite{Knox}-\cite{Lin},
in spite off, different models with dark energy can furnish the same age for an expanding universe.  However, considering 
the age of the universe at different eras and comparing with the age estimates of high-red shift objects, this degeneracy 
can be eliminated \cite{Friaca}.

The standard FRW model indicates a younger universe, if compared with estimates from globular cluster data \cite{Chaboyer}, 
and CMB measurements \cite{Spergel}.  
Using the expression for the Hubble function, given by the Eq.(12), and the correspondent function for the standard model,
 given by $H_0=\frac{2}{3 t_0}$, we can construct a quotient  between these functions, and observe the values for the 
$\epsilon$ parameter that furnish an older universe than the established by the standard model. We find that any positive 
value for the $\epsilon$-parameter results an older  universe. For increasing $\epsilon$ we obtain an older universe.

Let's suppose that the universe has the critical density. Using the current value for the Hubble function found by WMAP 
\cite{Spergel1}, $H_0 = 73.4 ^{+2.8} _{-3.8}$ km/s/Mpc, we can estimate a range for $t_0$ taking into account the FRW 
standard model, explicitly $t_0= (8.6-9.4)$ Gyr.  On the other hand, the experimental data that predicts the age of the 
globular cluster indicates an interval (10.6-12.27) Gyr \cite{Chaboyer}. Consequently we have a problem with respect to
 the age of the universe.
Note that, the predict age for the universe around $\epsilon =0$ is 12.8 Gyr, minimal value for $t_0$ in this model. So,
 this model is not plagued with the age problem, since that the age of the globular cluster does not furnish an upper limit to $t_0$ .

An additional constraint can be obtained using the accelerated process of the universe expansion, that indicates a past 
deceleration ($q<0$) beyond the red-shift $z_t = 0.46 \pm 0.13$ at 99.8$ \% $ at confidence level, where the subscript 
$t$ refers to the transition point which the universe change the signal of the deceleration parameter.  Using the expression 
for the deceleration parameter (Eq.13), we can write an expression for the transition redshift, namely
\begin{equation}
z_t=\{\frac{1-\epsilon}{2}\frac{3\Omega_{dm0}}{3\Omega_{de0}-\epsilon} \}^{\frac{1}{\epsilon -3}}-1 \, ,
\end{equation}
where the profile appear in the (Fig.3).  For $\epsilon = 0.01$, we obtain the transition redshift around $z_t=0.68$,
considering $\Omega_{dm}=0.3$, and $\Omega_{de}=0.7$. The expressions (13) and (14) are growing functions
of the $\epsilon$-parameter.
\begin{figure}[!ht]
{\centerline {\includegraphics[width=6cm]{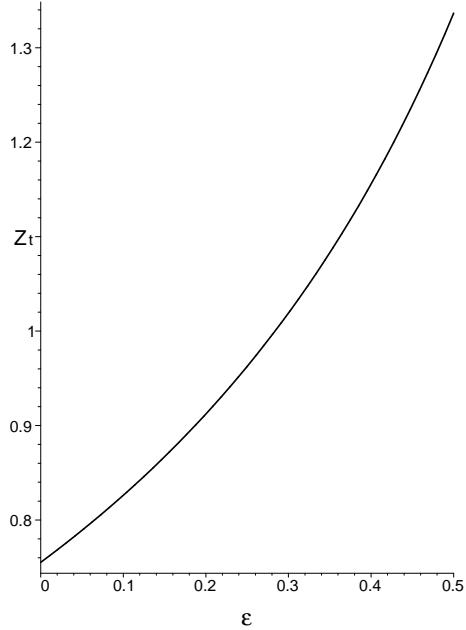}}}
\caption{Transition redshift as a function of  the $\epsilon$ parameter. With increasing of $\epsilon$ we obtain an 
early estimate for the transition redshift, and consequently, an older universe.}
\label{Fig.3}
\end{figure}
\section{Luminosity distance and acoustic scale}
Considerations about the accelerating expansion of the cosmos and the consequent existence of a dark component comes from geometrical tests that measures the Hubble expansion at various redshifts. One of then is the luminosity distance from standard candles.

The comoving distance $r(t,t_0)$ traveled by a light signal from a time $t$ to the present time is given by
\begin{equation}
r(t,t_0)=\int_t ^{t_0} \frac{d  t^{\prime}}{R(t^{\prime} )}\, ,
\end{equation}
considering  flat spatial sections. For redshift as an integration variable we have
\begin{equation}
r(z)=\int _0 ^z \frac{d z^{\prime}}{H(z^{\prime})}\, .
\end{equation}
With auxilious of the Eqs.(12), we can integrate  (16), resulting
\begin{equation}
r(z)=-\frac{H_0}{\Omega _{de0}}_2F_1\{[\frac{1}{2},\frac{1}{7-\epsilon}],[\frac{8-\epsilon}{7-\epsilon}],(1+z)^{7- 
\epsilon} \frac{3\Omega _{dm0}}{(3-\epsilon)\Omega _{de0}}\} \, ,
\end{equation}
where $_2F_1$ denotes a hypergeometric function. To illustrate the evolution of the comoving distance  for 
different values for the $\epsilon$-parameter, we show the  profile in the (Fig.4).
\begin{figure}[!ht]
{\centerline {\includegraphics[width=6cm]{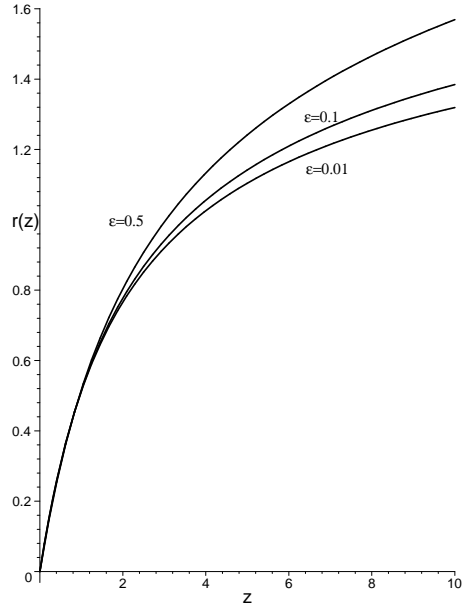}}}
\caption{ Profile for the luminosity distance versus redshift taking into account $\epsilon = 0.01\,, 0.1\,,0.5$. }
\label{Fig.4}
\end{figure}

Indeed, the modulus distance is given by the formula
\begin{equation}
 \mu(z)=5log(\frac{d_L}{Mpc})+25 \nonumber \, ,
\end{equation}
where the luminosity distance can be written as
\begin{equation}
 d_L=(1+z) \int _0 ^z \frac{du}{H(u)} \, . \nonumber
\end{equation}
In the Fig.5 we show the profile for the modulus distance versus redshift taking into account $\epsilon=0.5$
, and compare with the Union Sample of 557 Supernova Ia \cite{Ama} to illustrate the good agreement
 of the model with respect to the observational data. The profile for different values of the $\epsilon$-parameter
 do not help to decide about the best agreement with SnIa data.
\begin{figure}[!ht]
{\centerline {\includegraphics[width=6cm]{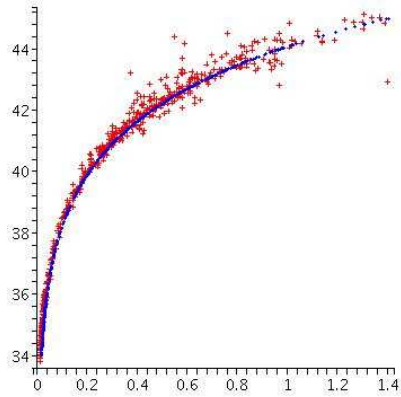}}}
\caption{ Profile for the modulus distance versus redshift taking into account $\epsilon=0.5$.
 The data of the Union Sample of 557 Supernova Ia are the red points and the theoretical results appear as blue points.
 We consider $\Omega_{de0}\approx 0.7$, $\Omega_{dm0}\approx 0.3 $ and the present value for the Hubble constant as $72 km/s/Mpc$.}
\label{Fig.4}
\end{figure}
Let us see this, performing a chi-square analysis using
\begin{equation}
 \chi ^2(\epsilon)=\sum_{i=1}^{557}\frac{[\mu(\epsilon; z_i)_{theoretical}-\mu(z_i)_{observational}]^2}{\sigma(z_i)^2}\, ,
\end{equation}
where $\sigma(z_i)$ is the correspondent 1$\sigma$ uncertainty.
The observational data are consistent with the considered model if $\frac{\chi ^2}{N-m}\leq 1$, where $N$ is 
the range of the data set used, and $m$ is the number of parameters. 

In the table I
\begin{figure*}[!ht]
\centerline{
\begin{tabular}{|c|c|c|c|c|c|}
\hline $\epsilon$ & 0.01 & 0.1 & 0.5 & 0.6 & 0.7  \\
\hline $\chi ^2$ &546.97&545.70&543.82&544.23&544.97 \\\hline
\end{tabular}}
{\centerline{Table I}}
\end{figure*}
we show that $\chi^2$ values that we find for different values for the $\epsilon$-parameter.  Note that for all values of 
$\epsilon$-parameter considered we have $\frac{\chi^2}{556} \leq 1$, that indicates the consistence of the model with the 
Supernova data, but the test is not conclusive in respect to the more adequate interval for the $\epsilon$-parameter. Although, 
the range $0.5 \leq \epsilon \leq 0.6 $ furnishes a better agreement.

On the other hand,with the universe expansion, the coupling reactions becomes inefficient.  Neutral atoms are formed and the ionization
 fraction freezes out.  The photons become free and the lack of further interactions preserves the density irregularities,
 imprinted on the photons field. Consequently, the density perturbations in the coupled baryon-photon fluid in the 
pre-recombination epoch are responsible by the dominant acoustic anisotropy in CMB.  Applying the classical angular 
diameter distance to CMB, we can learn about cosmological parameters by observing the anisotropy acoustic peak locations.  

The sound horizon scale is the maximum distance that a sound wave could have traveled in approximately 300.000 yrs 
from the beginning of the matter era until the time of recombination.  The angular diameter distance translate the $\Theta$ 
angle into a multipole $l$, or a length scale.  Therefore, one expects acoustic normal modes that are linked to the harmonic 
series of anisotropies.  

In order to obtain the multipole spacing for cosmological models we need of the angular  diameter distance, 
the sound horizon scale and the redshift at decoupling, that is the epoch when the physical imprint of the 
acoustic anisotropies in the CMB temperature pattern occurs, and the photon become unaffected by further 
interactions with the matter. The angular scale of the peaks of the angular power spectrum of the cosmic 
microwave background anisotropies is given by $\Theta _a = \frac{\pi}{l_a}$, where the multipole associated
 to the angular scale $\Theta _a$  is given by \cite{Page}
\begin{equation}
l_a =\pi \frac{r(z_{dec})}{r_s(z_{dec})} \, .
\end{equation}
The $r(z_{dec})$ is the comoving distance at decoupling, and $r_s(z_{dec})$ is the comoving size of the sound horizon at decoupling,
 that obeys \cite{Hu}
\begin{equation}
r_s(z_{dec})= \int ^{\frac{1}{1+z_{dec}}} _{0}  \frac{\overline{C}_s(R) dR}{R^2 H(R)} \, ,
\end{equation}
where the average sound speed before last scattering is given by $\overline{C}_a (a)= \frac{1}
{\sqrt{3+\frac{9\Omega_{b}}{4\Omega_\gamma}R}}$, and $\Omega_{b}$, $\Omega_\gamma$ 
are the ratio for baryons and radiation, respectively.

The component of the dark energy can be taken negligible in the calculus of the sound 
horizon, and numerical simulations indicates an error of order of $10^{-5} \% $ , resulting \cite{Hu}
\begin{equation}
r_s=\frac{4}{3H_0}\frac{\Omega _\gamma}{\Omega _{dm0}\Omega _b }\ln{\frac{[1+A_{dec}]^{1/2}+[A_{dec}+A_{eq}]^{1/2}}{1+A_{eq}^{1/2}}}\, ,
\end{equation}
where $A=\frac{3\Omega_b}{4\Omega_\gamma}R$.

With help of  Eqs. (17) and (20) we can show the profile for the multipole associated to the angular scale $\Theta_a$ , 
as function of the $\epsilon$ parameter (Fig.5).  We consider the decoupling redshift $z_{dec}=1089$ and the acoustic 
scale as $300 \pm 3$ \cite{Page}. So, we can infer an interval for the $\epsilon$ parameter, namely $\epsilon = 0.58-0.60$.
\begin{figure}[!ht]
{\centerline{\includegraphics[width=6cm]{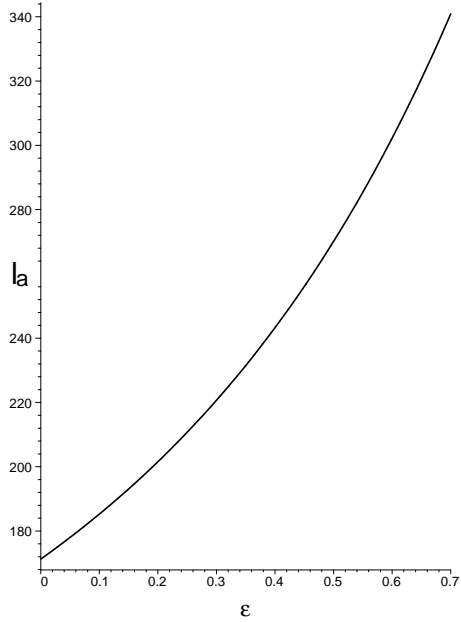}}}
\caption{Acoustic scale as function of the $\epsilon$-parameter.}
\label{Fig.5}
\end{figure}

The baryon acoustic oscillations(BAO) occurs at relatively large scales, but acoustic signature has been
detected at low redshift using 2dF Galaxy Redshift Survey \cite{Coles}, and the Sloan Digital Sky Survey \cite{York},
estimating the distance-redshift relation at $z=0.2$ and $z=0.35$, respectively. The observed scale of the BAO calculated from
these samples are analyzed and used to constrain the form of the distance
\begin{equation}
 D_v(z)=[(1+z)^2D^2_a(z)\frac{z}{H(z)}]^{\frac{1}{3}} \, ,
\end{equation}
where $D_A(z)=\frac{d_L(z)}{(1+z)^2}$ is the proper angular diameter distance, and $d_L(z)$ is the luminosity distance.

Matching the BAO to acquire the same measured scale at all redshifts we have \cite{Percival}, \cite{Saridakis}
\begin{equation}
 \chi^2_{BAO}=\frac{[\frac{D_V(0.35)}{D_V(0.2)}-1736]^2}{0.065^2} \, ,
\end{equation}
that allow us to find the reduced $\chi ^2_{BAO}$ for different $\epsilon$ values (Table II). \\
\begin{figure}[!ht]
\centerline{
\begin{tabular}{|c|c|c|c|c|c|}
\hline $\epsilon$ & 0.01 & 0.1 & 0.5 & 0.6 & 0.7  \\
\hline $\chi_{BAO} ^2$ &1.27&1.14&0.9&0.79&0.79 \\\hline
\end{tabular}}
{\centerline{Table II: Reduced $\chi ^2 _{BAO}$.}}
\end{figure}
Note in the table II that the BAO at low redshifts indicates $\epsilon > 0.5 $, in concordance with the BAO at large redshifts.

The validity interval that we find for the $\epsilon$-parameter states how essential is the coupling between the dark components, 
and also indicates that the addition of a cosmological constant, possibly cannot describe the dynamics of the universe. 

In several opportunities in the literature, the $\Lambda$-CDM model appear with good agreement to the observational data, and by
several authors the model is considered as a paradigm. In the analyses of the different models, the statefinder parameters
, introduced by Shani, Saini and Starobinsky \cite{Shani}, furnishes an qualitative idea, an geometrical diagnostic \cite{Alam},\cite{Zhang}
 of  how much the considered model is ``distant'' of the $\Lambda$-CDM. The statefinder pair is defined by:
\begin{eqnarray}
 r&=&\frac{d^3 R/dt^3}{(RH^3)} \, , \\
 s&=&\frac{r-1}{3(q-\frac{1}{2})} \, .
\end{eqnarray}

For LCDM (Lambda cold dark matter model) model $\{ r,s \}
=\{1,0 \}$ is a fixed point and the departure from this
point increase the distance from flat LCDM model \cite{Alam}. 

It is not difficult express the statefinder parameter $r$ in terms of the deceleration parameter and your redshift derivative, namely
\begin{equation}
r=2q^2 + q +q^{\prime}(1+z)\, .
\end{equation}
Consequently, the expressions for the statefinder parameters $r(t)$ and $s(t)$ are, respectively:
\begin{eqnarray}
 r(t)&=&1+\frac{\epsilon^2-3\epsilon}{2\{\cosh{\sqrt{KI}(\epsilon -3)t/2}\}^2+(\epsilon -3)}\\
 s(t)&=&\frac{(3-\epsilon)\epsilon}{3\{\cosh{\sqrt{KI}(\epsilon -3)t/2}\}^2}
\end{eqnarray}
 
In the Fig.(7) we display the  profile $s \times r$, and we note the sensibility of the statefinder parameters 
for relatively close values for the $\epsilon$-parameter. Note that for high redshift the statefinder parameters of the 
interacting model that we study is close to point \{1,0\}, characteristic for the LCDM model.
\begin{figure}[!ht]
{\centerline{\includegraphics[width=6cm]{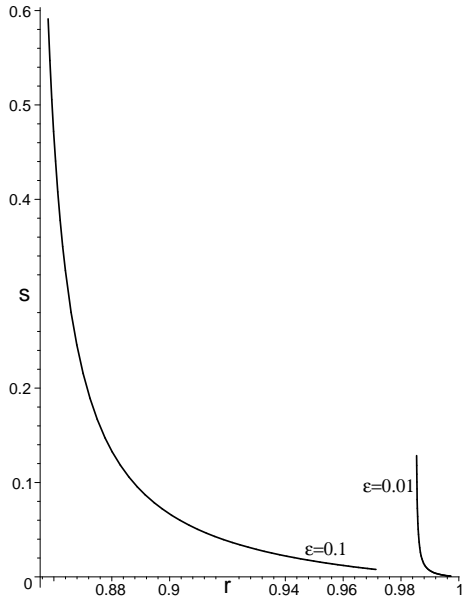}}}
\caption{Time evolution of the statefinder pair \{r,s\} for $\epsilon=0.01$ and $\epsilon=0.1$.The direction
for the increasing redshift is identical to the direction of the increasing $\epsilon$-parameter.}
\label{ Fig.6}
\end{figure}
\section{Conclusions}
In this work we consider a coupling between the dark components of the universe, where the dark matter density will dilute 
in a different rate.  So, we have creation of dark matter particles at expenses of dark energy, and the $\epsilon$-parameter
is linked to the creation process.
An interesting feature of the models with only one parameter is related to the coincidence problem, that contrary to 
a cosmological principle stating that we are in a special era of the universe. With only one parameter governing the dynamics
the coincidence problem is alleviate, and in some sense eliminated .
Taking into account the Union Sample of 557 Supernova Ia, 2dF Galaxy Redshift Survey and the Sloan Digital Sky Survey, we find
$0.5 \leq \epsilon \leq 0.6$, considering the first experiment, and $\epsilon \geq 0.5$ for the second and third experiments.
The interval obtained in the cited experiments do not contradict the previsions for the age of the universe, but 
the transition redshift obtained have a highest value than the established in the present literature.
\section*{Acknowledgements}
The author acknowledges the suggestions of the anonymous referee to the paper.

\end{document}